\documentclass[a4paper,fleqn]{cas-dc}

\usepackage{hyperref}
\usepackage[numbers]{natbib}

\def\tsc#1{\csdef{#1}{\textsc{\lowercase{#1}}\xspace}}
\tsc{WGM}
\tsc{QE}

\begin{document}
\let\WriteBookmarks\relax
\def\floatpagepagefraction{1}
\def\textpagefraction{.001}

\shorttitle{Fingerprints of Mott-Hubbard physics in LaTiO$_3$}    

\shortauthors{N.N. Kovaleva}  

\title[mode = title]{Fingerprints of Mott-Hubbard physics in optical spectra of antiferromagnetic LaTiO$_3$}  



%

\author[1]{N.N. Kovaleva}[orcid=0000-0002-4970-7782]

\cormark[1]





\credit{Conceptualization of this study, Investigation, Project administration, Writing}

\affiliation[1]{organization={P.N. Lebedev Physical Institute of the Russian Academy of Sciences},
            addressline={Leninsky prospekt, 53}, 
            city={Moscow},
            postcode={119991}, 
            country={Russia}}

\cortext[1]{Corresponding author}



\begin{abstract}
Magnetic properties of Mott-Hubbard insulators are determined by superexchange interactions mediated by the high-spin (HS) and low-spin (LS) intersite $d$--$d$ charge excitations, which can be associated with the HS- and LS-Hubbard subbands in optical experiments. To explore the Mott-Hubbard physics in orthorhombic LaTiO$_3$ crystal exhibiting the G-type antiferromagnetic order at the N\'{e}el temperature $T_{\rm N}$\,=\,146\,K, we present a comprehensive spectroscopic ellipsometry study in the spectral range 0.5--5.6\,eV at temperatures 10\,K\,$\leqslant$\,T\,$\leqslant$\,300\,K. We found that the complex dielectric function spectra of LaTiO$_3$ crystal are almost featureless, nearly isotropic, and weakly temperature dependent in the range of $d$--$d$ optical transitions. Nonetheless, analyzing the difference spectra below the $T_{\rm N}$, we have identified the LS-state $d^1d^1$\,$\rightarrow$\,$d^2d^0$ excitations at $\sim$\,3.7 and $\sim$\,5.15\,eV and estimated values of the on-site Coulomb repulsion $U$\,$\sim$\,4.2\,eV and Hund's exchange constant $J_H$\,$\sim$\,0.5\,eV, which define the energy of the HS-state $d^1d^1$\,$\rightarrow$\,$d^2d^0$ excitation at $\sim$\,2.7\,eV. In addition, we discovered that the pronounced lowest-energy 1.3\,eV optical band displays the critical intensity behavior and anomalous broadening with decreasing temperature below the $T_{\rm N}$. The discovered properties indicate that the 1.3\,eV band in LaTiO$_3$ can be associated with a Mott-Hubbard exciton.
\end{abstract}


\maketitle
\section{Introduction}\label{1}
\begin{figure}[tbp]
\includegraphics*[width=85mm]{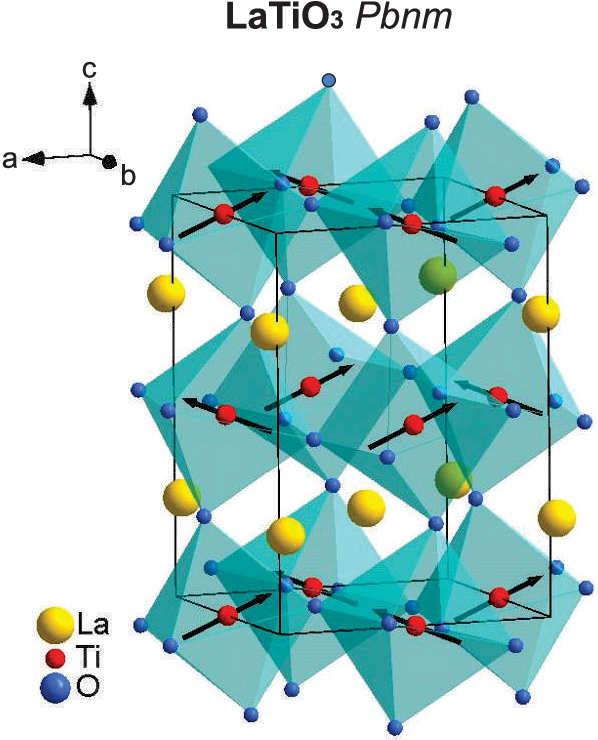}
\caption{Orthorhombic structure of LaTiO$_3$ crystal (space group $Pbnm$) with schematically shown G-type AFM alignment of Ti spins.}
\vspace{-0.3cm}
\label{fig1}
\end{figure}

Transition metal oxides demonstrate a rich variety of electronic, lattice, and magnetic properties, including colossal magnetoresistance (CMR) and high-temperature superconductivity (HTSC) 
\cite{Griffith,Tokura,Dagotto,Tajima,PWAnderson,KugelKhomskii,Keimer,
Bernhard,Knafo,Kovaleva_LMO_PRL,Kovaleva_LMO_PRB,Kovaleva_PRB_YTO,
Kovaleva_JETP_2016,Kovaleva_PLA_YTO,Kovaleva_PLA_LTO}. Among these, rare-earth ($R$) $R$TiO$_3$ titanates that crystallize in the orthorhombically distorted perovskite structure (of $Pbnm$ symmetry) are Mott-Hubbard insulators in which spin, charge, and orbital degrees of freedom are intrinsically entangled. In particular, $R$TiO$_3$ titanates amuse by the paradigm where the antiferromagnetic (AFM) G-type ground state for $R$\,=\,La,...,\,and\,Sm with the magnetic moment aligned parallel to the $a$ axis (see schematic presentation in Fig.\,\ref{fig1}) turns to the ferromagnetic (FM) ground state for $R$\,=\,Gd,...,Yb,\,and\,Y with the magnetic moment preferentially aligned parallel to the $c$ axis. For the $R$TiO$_3$ series, the most exemplary cases are provided by the end-point members, LaTiO$_3$ and YTiO$_3$: LaTiO$_3$ is an AFM with the N\'eel temperature $T_{\rm N}$\,=\,150\,K \cite{Cwik} and YTiO$_3$ is a FM with a comparatively low Curie temperature $T_{\rm C}$ of 30\,K \cite{Garrett}.

The magnetic properties of Mott-Hubbard insulators are determined by the electronic superexchange (SE) interactions \cite{PWAnderson,KugelKhomskii}, including both spin and orbital degrees of freedom, which are associated with the low-energy virtual charge excitations between two neighboring transition metal ions occurring via oxygen ligand orbitals, $d_i^md_j^m \rightleftharpoons d_i^{m+1}d_j^{m-1}$. The high-spin (HS) and low-spin (LS) contributions to the SE follow from the multiplet structure of the excited transition metal ion \cite{TanabeSugano} which, in turn, depends on the parameters of the on-site Coulomb repulsion $U$ and Hund's exchange constant $J_H$\cite{Oles1,Oles}. In LaMnO$_3$, a representative famous compound of CMR manganites, strong electron-lattice Jahn-Teller (JT) instability for the $d^4$ Mn$^{3+}$ ion removes the twofold $e_g$ orbital degeneracy on behalf of the MnO$_6$ octahedra distortions 
\cite{JahnTeller,OpikPryce,Stoneham,BVO,KaplanVekhter,Bersuker,Kovaleva_JPCM_LMO}. It was demonstrated by our comprehensive spectroscopic ellipsometry study of LaMnO$_3$ that the energies of $d_i^4d_j^4 \rightleftharpoons d_i^5d_j^3$ optical excitations by the $e_g$ electrons and variation of the associated optical spectral weight (SW) of the Hubbard subbands around the N\'{e}el temperature $T_{\rm N}$=140\,K provide an important information on the spin-orbital physics and even magnetic exchange constants in this compound \cite{Kovaleva_LMO_PRL,Kovaleva_LMO_PRB}.

In $R$TiO$_3$ crystals, the ground state of the $d^1$ Ti$^{3+}$ ion can be regarded as one electron in three $t_{2g}$ orbitals, where the threefold $t_{2g}$ degeneracy can partially be removed by the JT mechanism \cite{Mochizuki,Mochizuki1} or totally removed by the crystal-field (CF) effect driven by $R$\,--\,O covalency \cite{Pavarini,Pavarini1}. The excited states for the $d^2$ Ti$^{2+}$ ion are generated by the intersite $d_i^1d_j^1 \rightleftharpoons d_i^2d_j^0$ electron transitions. It seems that optical SW of these transitions must vanish, however, a direct electric-dipole-allowed optical excitation is expected to occur via an oxygen $2p$-band excited state, which is strongly dipole-allowed. In early optical measurements of LaTiO$_3$ and YTiO$_3$, the optical conductivity at low photon energies was found to be very similar, the only difference was interpreted as the rigid shift of the Mott gap to low photon energies in LaTiO$_3$ \cite{Arima}. However, the $d$\,--\,$d$ excitations associated with different Hubbard subbands were not separated from each other because of the practically featureless low-energy spectra. The spectral location of the intersite $d$\,--\,$d$ excitations, as well as the associated optical SW, can be monitored in the optical response due to the onset of long-range spin-spin correlations around the associated magnetic transition temperatures. In our prior spectroscopic ellipsometry study of LaMnO$_3$ tracing the optical SW redistribution around the $T_{\rm N}$=140\,K, we were able to identify the HS- and LS-state transitions for the $d^5$ Mn$^{2+}$ ion, as well as to estimate the parameters of the effective on-site Coulomb repulsion $U$ and Hund's exchange constant $J_H$ \cite{Kovaleva_LMO_PRL,Kovaleva_LMO_PRB}.
Following the same approach in our subsequent optical experiments carried out on a well characterized YTiO$_3$ single crystal with $T_C$\,=\,30\,K (amongst the highest values reported for YTiO$_3$ \cite{Garrett}), we were able to resolve a weak absorption feature around 2.85\,eV which revealed an SW increase with decreasing temperature below $T_C$ \cite{Kovaleva_PRB_YTO}. Therefore, the latter was assigned to a HS ($^3T_1$) transition, where the electron is transferred with a parallel spin to an unoccupied $t_{2g}$ orbital on the neighboring Ti site resulting in the redistribution of the optical SW due to FM ordering below $T_C$. However, as to our knowledge, the $d$\,--\,$d$ excitations associated with different Hubbard subbands, as well as the optical SW redistribution between them induced by the G-type AFM spin ordering at $T_{\rm N}$=150\,K in LaTiO$_3$ crystals, were not investigated so far.

Moreover, an interesting physics to be explored in LaTiO$_3$ titanate is concerned with a Mott-Hubbard exciton (MHE). Recent optical studies monitoring changes in the associated SW revealed the evidence of an MHE in LaVO$_3$ vanadate, where the lowest-energy peak at 1.8\,eV was attributed to an excitonic resonance (see Ref.\,\cite{Lovinger} and references therein) certifying that the excitonic SW is strongly influenced by the presence of spin and orbital order. In addition, the optical bands arising above the absorption edge were considered as reasonable candidates for an MHE in some $R$TiO$_3$ titanates \cite{Gruninger}. In particular, the lowest-energy optical band appears in YTiO$_3$ at 1.9\,eV, being notably shifted from the HS $^3T_1$ state to lower photon energies \cite{Kovaleva_PRB_YTO}. However, the contribution from polaronic and polaron-excitonic optical bands cannot be excluded at these energies \cite{Gruninger,Kovaleva_JETP_2002,Kovaleva_PhysB,Kovaleva_PhysB_1}. Indeed, the modeling of polaron-related features in complex oxides such as CMR materials revealed that the total lattice relaxation energy (electronic plus ionic) for a hole self-trapped at the O site in LaMnO$_3$ is estimated to be quite large: about 2.4\,eV \cite{Kovaleva_JETP_2002,Kovaleva_PhysB}. Moreover, the above theoretical analysis suggests that the band near 2.3\,eV is rather associated with the presence of self-trapped O$^{-}$ holes in the nonstoichiometric transition metal oxide compound. In addition, the indirect $d$\,--\,$d$ optical transitions and/or weakly dipole-allowed CT transitions \cite{Zenkov} could contribute at low photon energies.

At present, fundamental questions relevant to the physics of an MHE in Mott insulators remain unanswered. In particular, the theoretical investigation of the line shapes of the exciton bandwidth in a Mott-Hubbard insulator presented in the studies \cite{Fisher,Brinkman,Nagaoka,Nagaoka1} can serve as a key experimental probe of the origin of the low-energy optical bands in LaTiO$_3$ and YTiO$_3$. An intrinsic broadening of the MHE bandwidth is predicted within an extended Hubbard model induced by the presence of an AFM spin order. This effect arises from limitations put on the migration of $d$-state holes by means of electron exchange. In the purely FM spin configuration, all potential walks are allowed, therefore, the exciton bandwidth can be quite narrow even for a broad $d$-hole band. By contrast, in the AF phase the hole propagation is maximally limited and, hence, the zone-center exciton can significantly be shifted and broadened into a wide band \cite{Fisher}. We believe that direct optical measurements comparing line shapes of the low-energy optical bands above and below $T_{\rm N}$ or $T_{\rm C}$ could shed light on their origin in LaTiO$_3$ and YTiO$_3$.

\section{Materials and Methods}\label{2}
The LaTiO$_3$ single crystals were grown by the standard solid-state reaction, and the final stage of growing was performed in a floating-zone image furnace (for more details see Ref.\,\cite{Cwik}). Special concern was given to the stoichiometry represented by an excess of oxygen in the LaTiO$_{3+\delta}$ formula. The stoichiometry of the grown crystals was finally attested by determining the N\'eel temperature in a SQUID magnetometer. For the nearly stoichiometric LaTiO$_3$ samples, $T_N$\,=\,146\,K was obtained \cite{Cwik} among the highest values reported so far for LaTiO$_3$. The grown crystals exhibited mosaic patterns due to twinning. The specific pattern and the size of the twin domains depend on local conditions determined by mechanical stresses and temperature gradients experienced during the crystal growth. Assuming the cubic perovskite structure, the as grown LaTiO$_3$ single crystal was first aligned along the $Pbnm$ orthorhombic $c$-axis direction using x-ray Laue patterns. Then the crystal was cut parallel to the plane formed by the cubic $\langle$110$\rangle$ direction and the orthorhombic $c$ axis. As a result, we were able to prepare the $bc$ plane of the grown LaTiO$_3$ crystal. For optical measurements, the $bc$-plane crystal surface was polished to optical grade. The ellipsometry measurements in the photon energy range 0.5--5.6\,eV were performed using a home-built ellipsometer of a rotating-analyzer type. The sample was fixed on the cold finger of a helium flow ultrahigh vacuum cryostat, which was preliminary evacuated to a base pressure of 5\,$\times$\,10$^{-9}$ Torr at room temperature.

\section{Results and Discussion}\label{3}
\begin{figure}[tbp]
\includegraphics*[width=85mm]{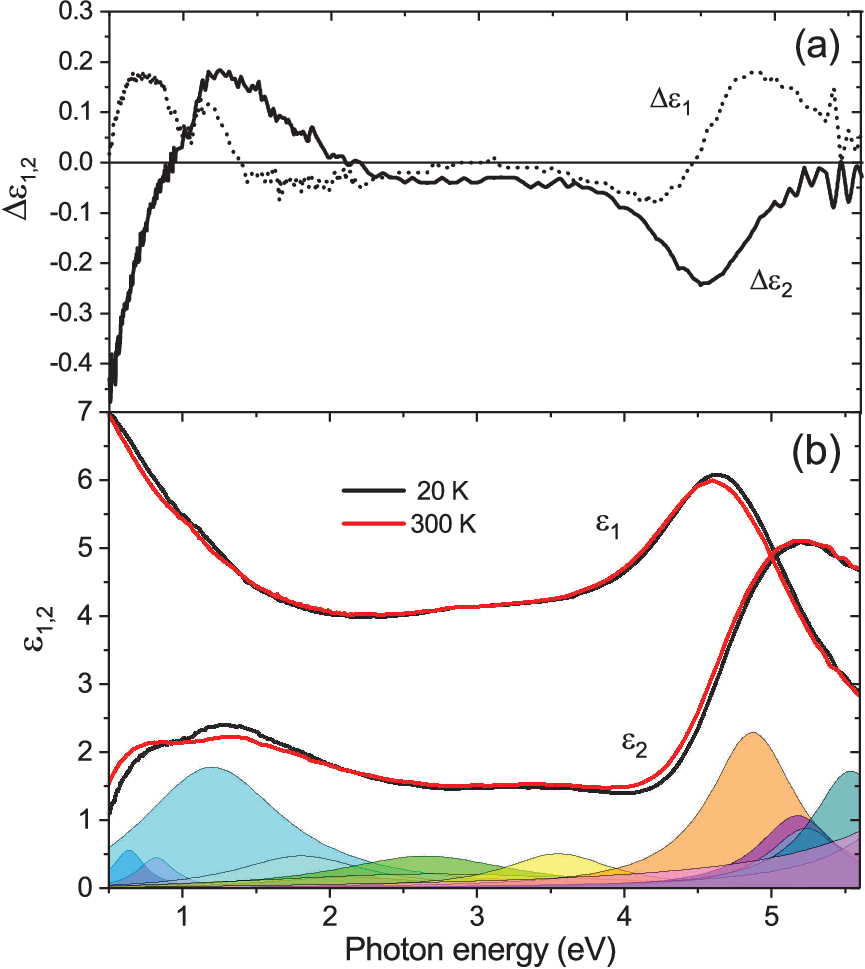}
\caption{(a) Difference spectra, 
$\Delta\varepsilon_1(\nu)=\varepsilon_1(\nu,20\,{\rm K})-\varepsilon_1(\nu,300\,{\rm K})$ and 
$\Delta\varepsilon_2(\nu)=\varepsilon_2(\nu,20\,{\rm K})-\varepsilon_2(\nu,300\,{\rm K})$, corresponding to (b) the representative low- and high-temperature spectra of the real 
$\varepsilon_1(\nu)$ and imaginary $\varepsilon_2(\nu)$ parts of the complex dielectric function spectra in LaTiO$_3$. The separate contributions of the Lorentzian bands resulting from the dispersion analysis of the low-$T$ (20\,K) spectra are clearly demonstrated. The corresponding parameters of the peak energies $\nu_j$, FWHM $\gamma_j$, and oscillator strength $S_j$ of the Lorentzian bands are listed in Table\,\ref{tbl1}.}
\vspace{-0.3cm}
\label{fig2}
\end{figure}
Optical spectra of a LaTiO$_3$ crystal in the investigated spectral range can have main contributions from (i) strong dipole-allowed $p$--$d$ charge-transfer (CT) transitions from the O($2p$) states in the valence band into the empty Ti($3d$) states in the conduction band, (ii) intrasite crystal-field (CF) $d$--$d$ transitions, and (iii) intersite $d$--$d$ transitions (progenitors of SE interactions). In addition, above the absorption edge one can anticipate contributions from polaron-excitonic bands \cite{Kovaleva_JETP_2002,Gruninger}, as well as from the indirect $d$--$d$ transitions. Many theoretical calculations \cite{Kovaleva_JETP_2002,Solovyev,Bouarab,Krasovska} agree that the optical band at $\sim$\,5.5\,eV, which is well pronounced in the optical spectra of manganites \cite{Kovaleva_LMO_PRL,Kovaleva_LMO_PRB} and titanates \cite{Kovaleva_PRB_YTO,Arima}, is associated with the strongly dipole-allowed $p$--$d$ CT transition (i). The intrasite CF $d$--$d$ transitions (ii) are associated with the electronic energies of a $d^1$ ion in a crystal, which are determined by the spin-orbital coupling and the energy due to the CF of the ligands \cite{Sturge}. Here, the gap between two cubic terms $\Delta_{CF}$=$E(^2E)-E(^2T_2)$ is predominant and can roughly be estimated as 2\,$\pm$\,0.5\,eV \cite{McClure,Iga,Higuchi1}. In addition, fine splitting occurs under trigonal CF and spin-orbit perturbations and the associated energy splittings can be recognized in infrared experiments (see Ref.\,\cite{Sturge} and references therein). Among the four $d^1_id^1_j\,\rightarrow\,d^2_id^0_j$ excitations (iii) there are the HS $^3T_1$ state at energy $U^*-3J_H$ and three LS states (those for the direct electric-dipole transition with spin flip), the doubly degenerate $^1T_2$ and $^1E$ states at energy $U^*-J_H$, and the $^1A_1$ state at energy $U^*+2J_H$ \cite{Oles}, which are determined by the values of the effective (screened) Coulomb interaction $U^*$ on the Ti ions and Hund's exchange $J_H$. According to the theoretical predictions \cite{Oles}, in the AF phase of LaTiO$_3$, all three Hubbard subbands should be evident in the optical spectra, the highest SW is associated with the doubly degenerate $^1T_2$ and $^1E$ states, and the SW of the HS $^3T_1$ state is expected to be relatively weak. 
\begin{table}[<options>]
\caption{Parameters $\nu_j$, $\gamma_j$, $S_j$, and $N_{eff}$ of Loretzians' peak energies, FWHM, dimensionless oscillator strength, and effective number of electronic charge carriers, respectively, obtained from the dispersion analysis of the complex dielectric function in $b$-axis ($bc$-plane) response of LaTiO$_3$ at $T$\,=\,20\,K, $\varepsilon_{\infty}$\,=\,1.62 (the parameters for $T$\,=\,300\,K are given in parenthesis, $\varepsilon_{\infty}$\,=\,1.56).}
\label{tbl1}
\begin{tabular*}{\tblwidth}{@{}CCCC@{}}
\toprule
\,\,\,\,\,$\nu_j$,\,eV&\,\,\,\,\,$\gamma_j$,\,eV&\hspace{0.6cm}$S_j$&N$_{eff}$\\ \midrule
0.66(0.59)&0.25(0.37)&0.22(0.34)&\\
0.84(0.77)&0.30(0.38)&0.16(0.25)&\\
1.34(1.30)&1.20(1.31)&1.53(1.54)&0.120\\
1.89(1.84)&1.07(1.12)&0.27(0.29)&\\
2.24(2.18)&1.31(1.34)&0.10(0.11)&\\
2.76(2.73)&1.48(1.56)&0.25(0.26)&0.083\\
3.10(3.11)&3.15(2.98)&0.19(0.21)&\\
3.12(3.11)&0.46(0.58)&0.02(0.03)&\\
3.59(3.59)&0.89(0.99)&0.13(0.13)&0.073\\
4.89(4.84)&0.72(0.74)&0.34(0.33)&\\
5.19(5.15)&0.65(0.63)&0.13(0.10)&0.153\\
5.25(5.19)&0.63(0.66)&0.11(0.13)&\\
5.25(5.50)&0.65(0.67)&0.20(0.20)&\\
\bottomrule
\end{tabular*}
\end{table}

Figure\,~\ref{fig2}(a,b) shows representative low- (20\,K) and high- (300\,K) temperature spectra of the real $\varepsilon_1(\nu)$ and imaginary $\varepsilon_2(\nu)$ parts of the complex dielectric function, $\tilde\varepsilon=\varepsilon_1$ + i$\varepsilon_2$, as well as the corresponding difference spectra, $\Delta\varepsilon_1(\nu)=\varepsilon_1(\nu,\,20\,\rm{K})-\varepsilon_1(\nu,\,300\,\rm{K})$ and $\Delta\varepsilon_2(\nu)=\varepsilon_2(\nu,\,20\,\rm{K})-\varepsilon_2(\nu,\,300\,\rm{K})$. In contrast to the YTiO$_3$ crystal, which exhibits a noticeable anisotropy for the $ab$-plane and $c$-axis response \cite{Kovaleva_PRB_YTO}, the complex dielectric function spectra of LaTiO$_3$ crystal are almost isotropic on the $bc$ plane. In virtue of the Mott-Hubbard nature of the optical gap in LaTiO$_3$ and YTiO$_3$, the absorption edge in these compounds is assumed to be of a $d$--$d$ origin. However, the assignment of the low-energy optical transitions in these compounds has not been clarified yet. Both YTiO$_3$ and LaTiO$_3$ exhibit the dielectric function spectra dominated by a pronounced lowest-energy optical band, which appears at $\sim$\,1.9\,eV (2.1\,eV) in the $b$-axis ($c$-axis) response in YTiO$_3$ \cite{Kovaleva_PRB_YTO}, being, however, significantly shifted to a lower energy of $\sim$\,1.3\,eV in LaTiO$_3$. The optical bands at higher energies form practically plain background for the intense optical band peaking at 5.2--5.7\,eV, which can be associated with the strongly dipole-allowed O($2p$)\,--\,Ti($3d$) CT transitions (i). From the dielectric function spectra presented in Fig.\,\ref{fig2}(b) one can see that their temperature changes are rather weak in the whole investigated temperature range from 300 to 20\,K. The most pronounced temperature changes are associated with the low-energy band situated above the absorption edge at $\sim$\,1.3\,eV. We note that, in contrast to the nearly temperature-independent $\varepsilon_2$ peak intensity of the 1.9\,eV optical band in YTiO$_3$ crystal \cite{Kovaleva_PRB_YTO}, the $\varepsilon_2$ peak intensity of the low-energy 1.3\,eV band notably increases in LaTiO$_3$ with decreasing temperature (see Fig.\,\ref{fig2} and Fig.\,12 of Ref.\,\cite {Kovaleva_PRB_YTO}). The details of the temperature dependence of the 1.3\,eV band will be further discussed.

To distinguish contributions from the separate optical bands, we applied the classical dispersion analysis in which the complex dielectric function is represented by a sum of Lorentz oscillators 
\begin{eqnarray}
\tilde\varepsilon(\nu)=\varepsilon_{\infty}+\sum_j\frac{S_j}{\nu_j^2-\nu^2-{\rm i}\nu\gamma_j}
\end{eqnarray}
due to contributions from different interband optical transitions, where the parameters $\nu_j$, $\gamma_j$, and $S_j$ specify the energy maximum, full width at half maximum (FWHM), and oscillator strength of the individual $j$th oscillator, respectively, and $\varepsilon_{\infty}$ is the dielectric function core contribution. In the classical dispersion analysis of the complex dielectric function, we input a minimal set of Lorenzians, which were fitted in the Kramers-Kronig consistent way to ensure the most reliable and stable parameters. The parameters $\nu_j$, $\gamma_j$, and $S_j$ of Lorentz bands obtained in the fitting of the low- and high-temperature complex dielectric function spectra presented in Fig.\,\ref{fig2}(b) are given in Table\,\ref{tbl1}. As an illustration, the Lorentzian bands for the 20\,K complex dielectric function spectra are clearly displayed in Fig.\,\ref{fig2}(b). The results of the dispersion analysis of the dielectric function spectra listed in Table\,\ref{tbl1} indicate that the wide-range temperature changes can predominantly be associated with ordinary lattice anharmonicity effects leading to narrowing of optical bands with decreasing temperature. The narrowing effect is clearly illustrated by the main optical band at around 5\,eV, which leads to a minimum in the difference spectra $\Delta\varepsilon_2(\nu)=\varepsilon_2(\nu,\,20\,\rm{K})-\varepsilon_2(\nu,\,300\,\rm{K})$ at $\sim$\,4.5\,eV, as can be seen from Fig.\,\ref{fig2}(a,b).
\begin{figure}[tbp]
\includegraphics*[width=85mm]{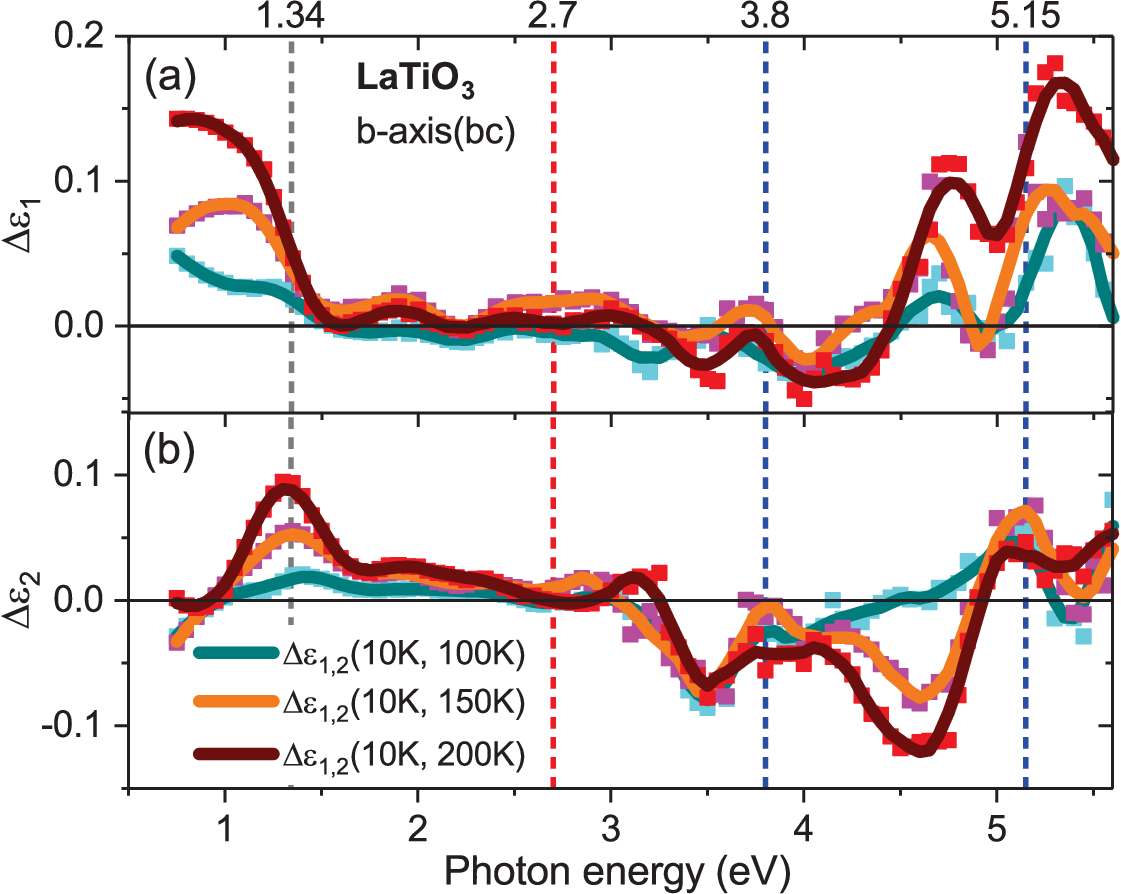}
\caption{Temperature changes in the difference spectra of (a) real $\Delta\varepsilon_1(10\,K,T)=\varepsilon_1(\nu,\,10\,K)-\varepsilon_1(\nu,\,T)$ and (b) imaginary $\Delta\varepsilon_2(10\,K,T)=\varepsilon_2(\nu,\,10\,K)-\varepsilon_2(\nu,\,T)$ parts of the complex dielectric function in the $b$-axis polarization. Red and blue dashed lines indicate tentative positions of the HS- and LS-state Hubbard bands, respectively.}
\vspace{-0.3cm}
\label{fig3}
\end{figure}
\begin{figure}[tbp]
\includegraphics*[width=85mm]{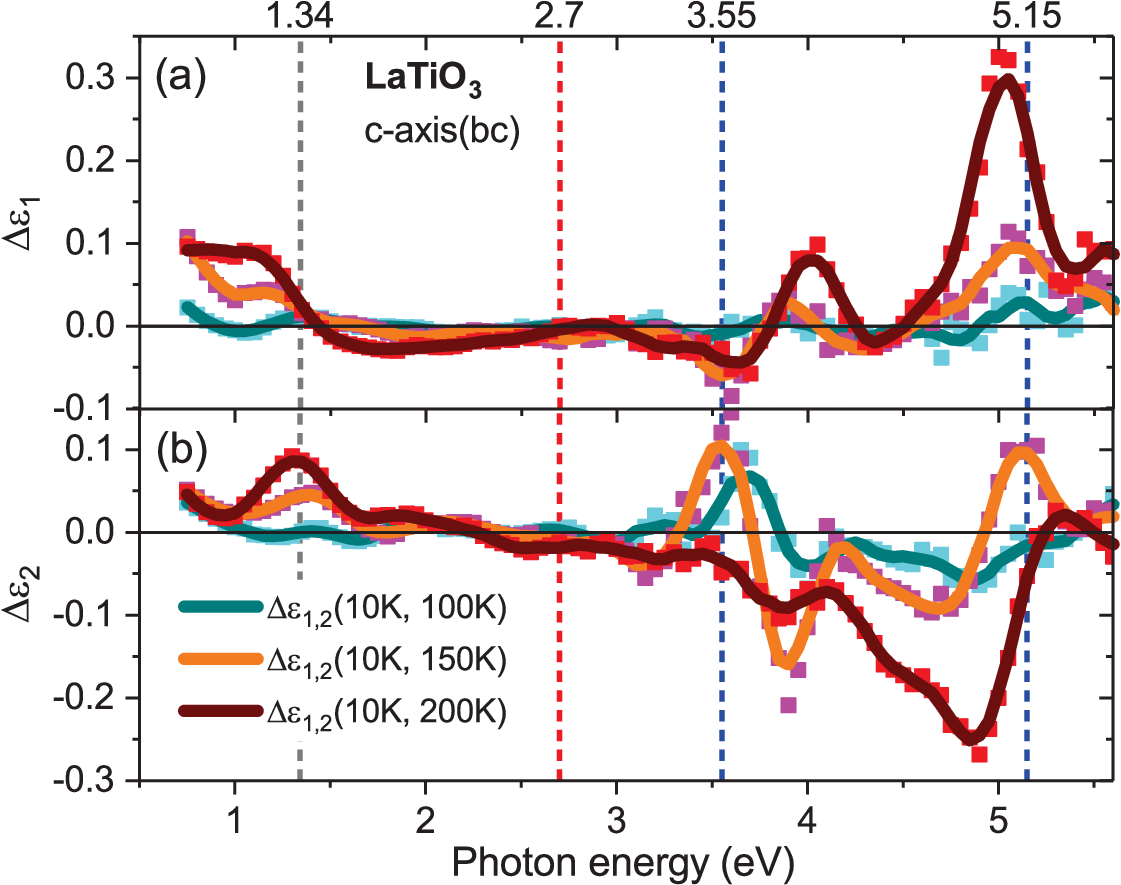}
\caption{Temperature changes in the difference spectra of (a) real $\Delta\varepsilon_1(10\,K,T)=\varepsilon_1(\nu,\,10\,K)-\varepsilon_1(\nu,\,T)$ and (b) imaginary $\Delta\varepsilon_2(10\,K,T)=\varepsilon_2(\nu,\,10\,K)-\varepsilon_2(\nu,\,T)$ parts of the complex dielectric function in the $c$-axis polarization. Red and blue dashed lines indicate tentative positions of the HS- and LS-state Hubbard bands, respectively.}
\vspace{-0.3cm}
\label{fig4}
\end{figure}

Meantime, our ellipsometry measurements allowed us to resolve much weaker temperature effects in the dielectric function spectra of LaTiO$_3$, which can be associated with a long-range spin ordering and redistribution of the optical SW around the $T_{\rm N}$\,=\,146\,K. Figures\,\ref{fig3} and \ref{fig4} demonstrate the difference spectra of the real and imaginary parts of the dielectric function, $\Delta\varepsilon_1(\nu)$ and $\Delta\varepsilon_2(\nu)$, between 10\,K and 100, 150, and 200\,K, obtained for the $b$- and $c$-axis response, respectively. From Fig.\,\ref{fig4} one can see that in the range of intersite $d$--$d$ transitions the $b$-axis difference spectra show $\Delta\varepsilon_2(\nu)$ resonance features at $\sim$\,3.8 and 5.15\,eV, where the highest intensity corresponds to the difference between 10\,K (for AFM ordered spins) and 150\,K (for fluctuating spins right above the $T_{\rm N}$) accompanied with $\Delta\varepsilon_1(\nu)$ antiresonance features. At similar energies of 3.6 and 5.15\,eV, the $\Delta\varepsilon_2(\nu)$ resonance features are present in the $c$-axis difference spectra (see Fig.\,\ref{fig4}). For comparison, the $\Delta\varepsilon_2(\nu)$ spectra between 10 and 200\,K display the pronounced deepening around 4.5\,eV with increasing temperature from 150 to 200\,K certifying an appreciable contribution from lattice anharmonicity effects (as it was discussed in the context of the data presented in Fig.\,\ref{fig2}), which leads to smoothing of the $d$\,--\,$d$ features associated with spin-spin correlations in the difference spectra.
\begin{figure}[tbp]
\includegraphics*[width=85mm]{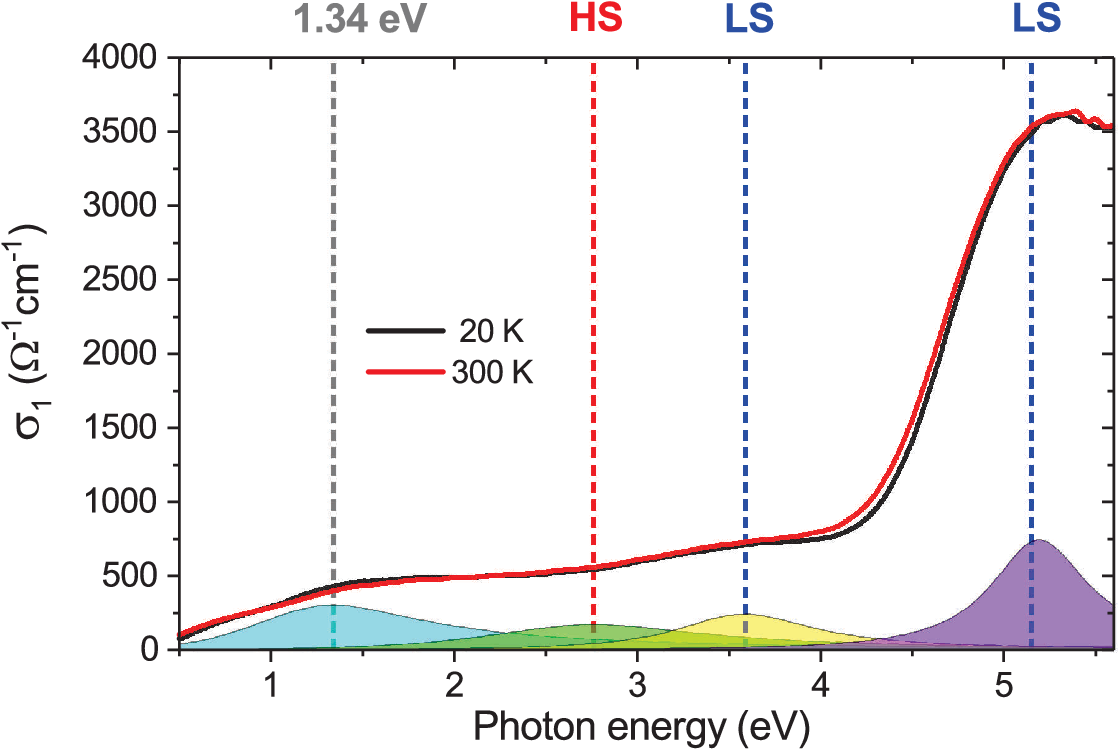}
\caption{Optical conductivity of LaTiO$_3$ single crystal in the $b$-axis response ($bc$ plane) at temperatures of 20 and 300\,K. The Lorentz bands associated with the 1.34\,eV lowest-energy band and the HS- and LS-state Hubbard subbands resulting from the dispersion analysis at 20\,K are displayed (see more details in Table\,\ref{tbl1}).}
\vspace{-0.3cm}
\label{fig5}
\end{figure}

The discovered optical bands at 3.7\,$\pm$\,0.1\,eV and 5.15 $\pm$ 0.1\,eV, which exhibit an increase in the $\varepsilon_2(\nu)$ with decreasing temperature below the $T_{\rm N}$, can be associated with the LS $d^1_id^1_j\,\rightarrow\,d^2_id^0_j$ transitions (iii) to two degenerate $^1T_2$ and $^1E$ states at energy $U^*-J_H$ and $^1A_1$ state at energy $U^*+2J_H$, where the electron is transferred to the $t_{2g}$ orbitals on the neighboring Ti site with an antiparallel spin. Then, from the system of equations, $U^*$ $-$ $J_H$ = 3.7 and $U^*$ $+$ $2J_H$ = 5.15\,eV, we can estimate the parameters $U^*$ = 4.2 $\pm$ 0.1\,eV and $J_H$ = 0.5 $\pm$ 0.1\,eV. The obtained $U^*$ value is well consistent with the effective on-site electron correlation energy $U_{dd}$\,$\sim$\,4\,eV estimated from the photoemission spectroscopy study in LaTiO$_3$ by Higuchi {\it et al.} \cite{Higuchi}.
In addition, there is a good agreement for the $U^*$ values resulting for the Ti oxides from x-ray emission spectroscopy and optical studies \cite{Kovaleva_PRB_YTO,Bocquet}, and theoretical $ab$ $initio$ values \cite{Aryasetiawan}. The obtained $J_H$ value is close to the free-ion value for the Hund's exchange coupling constant of 0.59\,eV \cite{Griffith}. Taking the obtained values of the parameters $U^*$ and $J_H$, we can propose the position of the HS $^3T_1$ transition with energy $U^*$\,$-$\,$3J_H$ at 2.7\,$\pm$\,0.1\,eV. According to the theory predictions, the HS-state transition associated with FM spin correlations is expected to be relatively weakly pronounced for G-type AFM ordering accomplished in LaTiO$_3$ \cite{Oles}. Indeed, we cannot see any clear evidence for the temperature dependence of the HS-state transition around 2.7\,eV (which is seemingly hidden within the experimental noise) in the difference spectra in Figs.\,\ref{fig3} and \ref{fig4}. In YTiO$_3$, the values of $U^*$\,=\,4.5\,$\pm$\,0.2\,eV and $J_H$\,=\,0.55\,$\pm$\,0.1\,eV obtained from our previous ellipsometry study \cite{Kovaleva_PRB_YTO} are similar to those obtained for LaTiO$_3$. However, in contrast to LaTiO$_3$, we observed that in YTiO$_3$ the optical band at $\sim$\,2.85\,eV exhibited an increase in the imaginary part of the dielectric function $\varepsilon_2(\nu)$ between 15 and 55\,K associated with FM spin ordering below the Curie temperature $T_C$\,=\,30\,K. Instead in YTiO$_3$, the LS-state transitions are appeared to be weakly pronounced making the assignment of the higher-energy LS-state transitions more difficult (see more details in Ref.\,\cite{Kovaleva_PRB_YTO}). In addition, we can recognize fingerpints of the HS-state transition in the $b$- and $c$-axis response represented by the minima in the $\Delta\varepsilon_2(\nu)$ at around 3.5 and 3.9\,eV, respectively (see Figs.\,\ref{fig3} and \ref{fig4}). These transitions can seemingly be associated with the $d^1_id^1_j\,\rightarrow\,d^2_id^0_j$ $^3T_2$ $t_{2g}$\,$\rightarrow$\,$e_g$ excitation, being shifted from the HS-state $^3T_1$ transitions by the CF gap of about 0.8\,-1.2\,eV. We note that the reduced value of the CF gap in LaTiO$_3$ can seemingly be associated with the $R$\,--\,O covalence effects \cite{Pavarini,Pavarini1,Kovaleva_PLA_LTO}.
\begin{figure}[tbp]
\includegraphics*[width=92mm]{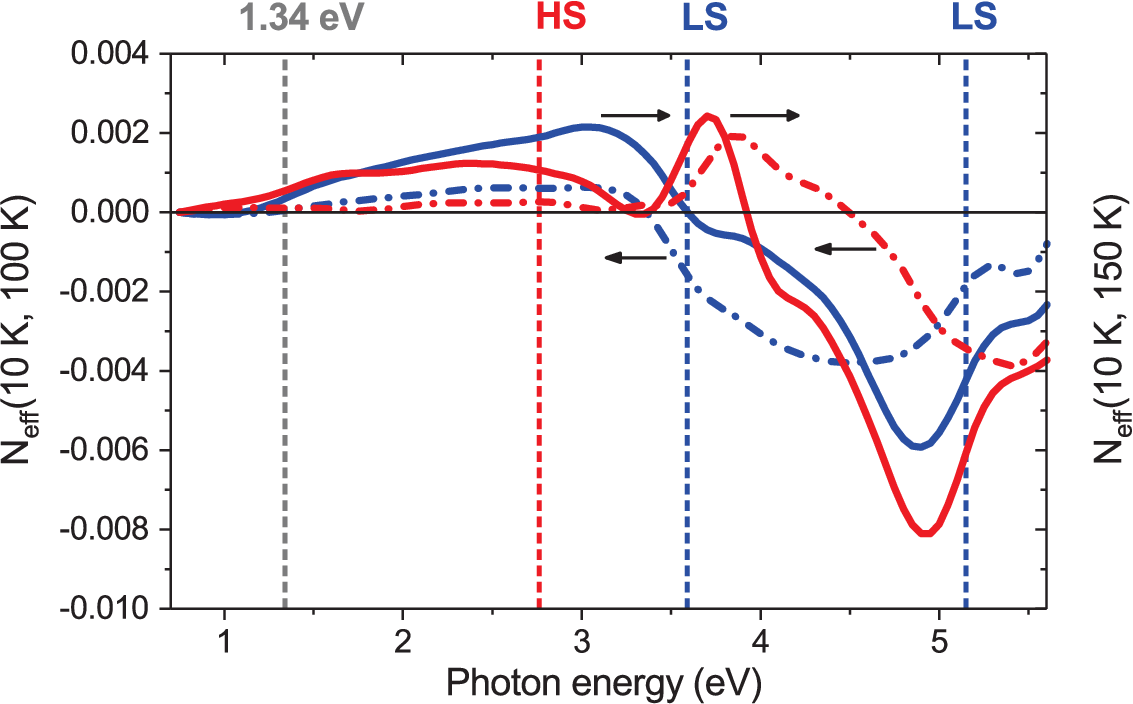}
\caption{Optical SW changes $\Delta N_{eff}(\nu)$ between 10\,K and 100\,K (dashed lines) and between 10\,K and 150\,K (solid lines) in the $b$- and $c$-axis response (shown by blue and red curves, respectively). The spectral positions of the lowest-energy 1.3\,eV band and the HS- and LS-state transitions are shown by vertical dashed lines.}
\vspace{-0.3cm}
\label{fig6}
\end{figure}

Figure\,\ref{fig5} shows the optical conductivity spectra $\sigma_1$ = $\frac{1}{4\pi}\nu\varepsilon_2(\nu)$ at temperatures of 20 and 300\,K obtained from the ellipsometry data presented in Fig.\,\ref{fig2}(b). In consistency with the prior work \cite{Arima}, the optical conductivity spectra show a smooth evolution of optical absorption with the onset at low photon energies below $\sim$\,0.5\,eV. According to our dispersion analysis for $T$\,=\,20\,K, the HS- (at $\sim$\,2.8\,eV) and LS- (at $\sim$\,3.6, and 5.2\,eV) Hubbard subbands are clearly displayed, and the lowest-energy optical band peaking at $\sim$\,1.3\,eV is presented in addition. An important characteristic of optical bands is the associated optical SW, which can be estimated for the separate Lorentz oscillator as SW\,=\,$\int \sigma_1(\nu')d\nu'$\,=\,$\frac{\pi}{120}S_j\nu_j^2$. The normalized SW values can be expressed in the effective number of electronic charge carriers $N_{eff}$\,=\,$\frac{2m}{\pi e^2N}$SW, where $N$\,=\,$a_0^{-3}$\,=\,1.7$\times$10$^{22}$\,cm$^{-3}$ is the density of Ti atoms and $m$ is the free electron mass. The associated $N_{eff}$ values for the 2.8\,eV HS-state of 0.08 and for the 3.6\,eV LS-state of 0.07 (see Table\,\ref{tbl1}) are very similar to those of 0.07 and 0.05, respectively, obtained from our previous optical study of YTiO$_3$ (see Table I of Ref.\,\cite{Kovaleva_PRB_YTO}). However, we would like to note that the low temperature $N_{eff}$ values for the lowest-energy band are notably different, being about 0.12 and 0.21 in LaTiO$_3$ and YTiO$_3$ crystals, respectively.

Figure\,\ref{fig6} demonstrates the SW changes $\Delta N_{eff}$(10\,K,\,100\,K) and 
$\Delta N_{eff}$(10\,K,\,150\,K) in the $b$- and $c$-axis response associated with the difference dielectric function spectra presented in Figs.\,\ref{fig3} and \ref{fig4}. The optical SW redistribution presented in Fig.\,\ref{fig6} indicates the transfer of the SW from above 3.5\,-4.5\,eV to lower photon energies as the temperature is lowered. Here, in the overall SW balance, not only the HS- and LS-Hubbard subbands are involved, but also the lowest-energy 1.3\,eV band and the HS-type transition peaking at $\sim$\,3.5 and $\sim$\,3.9\,eV in the $b$- and $c$-axis, respectively, of the $d^1_id^1_j$\,$\rightarrow$\,$d^2_id^0_j$ $^3T_2$ $t_{2g}$\,$\rightarrow$\,$e_g$ origin. One can notice that the SW decrease around the HS-state $^3T_1$ transition with decreasing temperature in the $c$ axis implies the presence of FM correlations, which can be associated with the weak FM moment available along the $c$ axis.
\begin{figure}[tbp]
\includegraphics*[width=80mm]{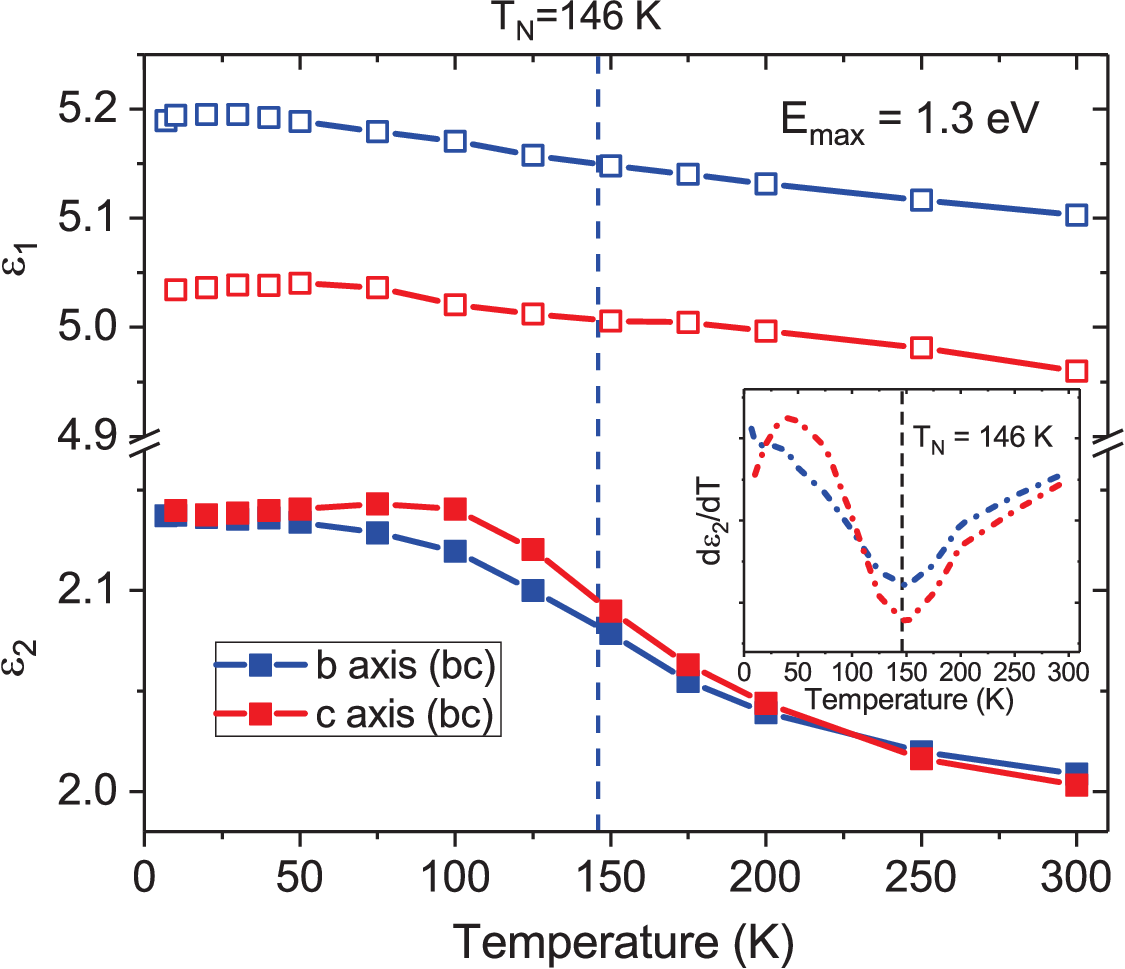}
\caption{Temperature variation of the $\varepsilon_1(\nu)$ and $\varepsilon_2(\nu)$ of the lowest-energy band peaking at 1.3\,eV in the $b$- and $c$-axis response. Inset: temperature dependence of the $d\varepsilon_2(\nu)/dT$.}
\vspace{-0.5cm}
\label{fig7}
\end{figure}
\begin{figure}[tbp]
\includegraphics*[width=80mm]{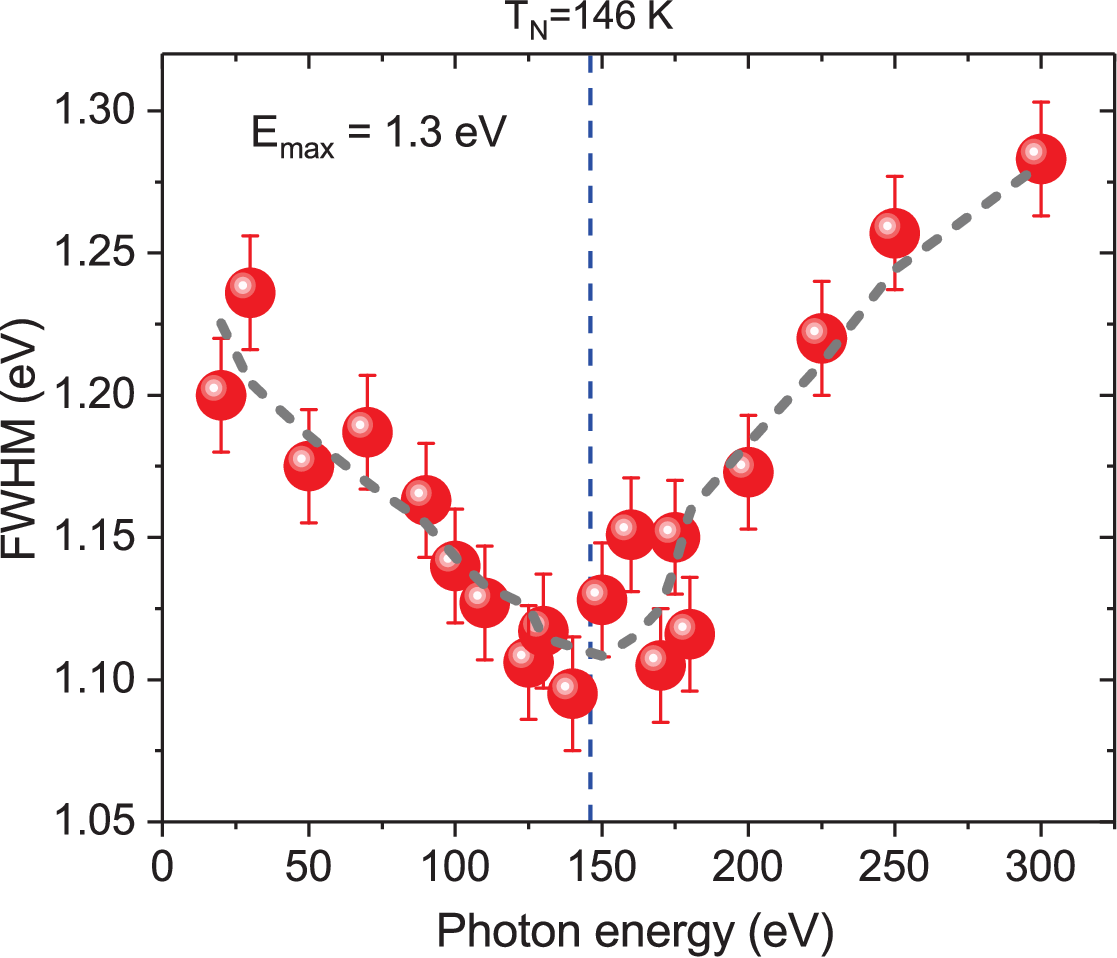}
\caption{Temperature variation of the FWHM of the lowest-energy band peaking at 1.3\,eV in the $b$-axis ($bc$ plane) spectra.}
\vspace{-0.5cm}
\label{fig8}
\end{figure}

Another interesting issue is associated with the nature of the lowest-energy optical band peaking at $\sim$\,1.3\,eV. First, in contrast to the nearly temperature-independent 
$\varepsilon_2$ peak intensity of the 1.9\,eV optical band in YTiO$_3$ crystal \cite{Kovaleva_PRB_YTO}, the $\varepsilon_2$ peak intensity of the 1.3\,eV low-energy band notably increases in LaTiO$_3$ with decreasing temperature (see Figs.\,\ref{fig3}, \ref{fig4}, and \ref{fig7}). Moreover, the derivative $d\varepsilon_2/dT$ shows a minimum near the AFM transition temperature $T_{\rm N}$\,=\,146\,K (see inset of Fig.\,\ref{fig7}). Second, from the results of the dispersion analysis of the temperature-dependent complex dielectric function spectra, we discovered that the FWHM of the 1.3\,eV band in LaTiO$_3$ shows a narrowing caused by decreasing lattice anharmonicity effects with decreasing temperature from 300\,K followed by the broadening below the N\'{e}el temperature 
$T_{\rm N}$\,=\,146\,K (see Fig.\,\ref{fig8}). The broadening is theoretically predicted \cite{Fisher} for a MHE, where the hole migration becomes limited in the presence of the AFM spin order. In addition, the spectral position of the lowest-energy band at $\sim$\,1.3\,eV is essentially shifted to lower photon energies in AFM LaTiO$_3$ compared to that at $\sim$\,1.9\,eV in FM YTiO$_3$. This can be consistent well with the theoretical predictions reported by Doniach {\it et al.} \cite{Fisher}. Indeed, an intrinsic broadening of the exciton bandwidth in a Mott-Hubbard insulator is predicted within an extended Hubbard model (in the strong repulsion limit $U \rightarrow \infty$) induced by the presence of spin order \cite{Fisher}. This phenomenon arises from restrictions put on the propagation of $d$-state holes by electron exchange. The only limitation for the migration of an exciton through the crystal arises through the $U$ repulsion exerting on the hole, the electron following the hole does not take part in the spin dynamics, and the exciton propagation is determined solely by the hole propagation. The theory of hole migration determined by the hole tunneling parameter $t$ was studied in the standard Hubbard model ($U \rightarrow \infty$) by using the moment method \cite{Brinkman,Nagaoka,Nagaoka1}. The $n$th moment of the spectral density is estimated being weighted over the sum of allowed walks on $n$ steps on the lattice accompanied by the electron spin rearrangement. The suitable walks include (i) removal of an electron from the chosen site generating a hole, (ii) moving the hole by exchanging with neighboring electrons carrying their spin, and (iii) in $n$ steps replacing the initially excited electron (with its spin) in the hole at some other site. As a result, the final spin configuration matches the initial one. The initial configuration is definite in the magnetically ordered phase (AFM) or can be purely random in the paramagnetic (PM) state. The spectral density moments are susceptible to the initial spin arrangement (AF or PM). Thus, a one step walk must involve a spin flip and, therefore, it is excluded in the AF phase but can occur with a $\frac{1}{2}$ probability in the PM phase. The line shapes have been presented in the study by Doniach {\t et al.} \cite{Fisher}. In the purely FM spin configuration, all potential walks are allowed and, therefore, the exciton bandwidth can be quite narrow even for a broad $d$-hole band. By contrast, in the AF phase the hole propagation is maximally limited and, as a result, the zone-center exciton could be significantly shifted and broadened into a wide band.

\section{Conclusion}\label{4}
The presented spectroscopic ellipsometry study reveals that the intrinsic low-energy complex dielectric function spectra of the LaTiO$_3$ crystal are almost featureless, nearly isotropic, and weakly temperature dependent. Even so, a comprehensive investigation of the difference in the dielectric function spectra with decreasing temperature below the AFM transition temperature $T_{\rm N}$\,=\,146\,K, in combination with a classical dispersion analysis allowed us to identify the $d^1d^1$\,$\rightarrow$\,$d^2d^0$ intersite electronic transitions associated with the LS electron excitations with spin flip: the doubly degenerate $^1T_2$ and $^1E$ states at 3.7\,$\pm$\,0.1\,eV and the $^1A_1$ state at 5.15\,$\pm$\,0.1\,eV, as well as to estimate the parameters of the effective on-site Coulomb repulsion $U^*$\,=\,4.2\,$\pm$\,0.1\,eV and Hund's exchange constant $J_H$\,=\,0.5\,$\pm$\,0.1\,eV. The intersite $d$--$d$ transitions appear in the optical response of LaTiO$_3$ and YTiO$_3$ below the onset of the strongly dipole-allowed $p$--$d$ transition at $\sim$\,5.2\,eV at similar energies. However, the HS $^3T_1$ transition at 2.7\,$\pm$\,0.1\,eV is {\it de facto} silent in the AFM phase of LaTiO$_3$, in contrast, it is weakly displayed at $\sim$\,2.85\,eV in the FM phase of YTiO$_3$ below the $T_C$\,=\,30\,K \cite{Kovaleva_PRB_YTO}. In addition, in the optical response of LaTiO$_3$ the pronounced lowest-energy band is present, whose origin, as to our knowledge, remained unclear. The spectral position of the lowest-energy band appearing in LaTiO$_3$ at 1.3\,eV is notably shifted to lower photon energies from that in YTiO$_3$ appearing at 1.9-2.1\,eV, and the associated optical SWs are also very different. In addition, the 1.3\,eV optical band in LaTiO$_3$ displays the critical intensity behavior and broadening with decreasing temperature below the $T_{\rm N}$. The observed broadening with decreasing temperature is highly unusual and leads to an increase in the optical SW, which can be associated with the kinetic energy enhancement in the course of the partial delocalization process caused by the onset of the AFM order in LaTiO$_3$ crystal. The observed properties of the lowest-energy band in LaTiO$_3$ and YTiO$_3$ are well consistent with the theoretical predictions reported for a Mott-Hubbard exciton affected by the AFM or FM spin order in Mott-Hubbard insulators by Doniach {\it et al.}  \cite{Fisher}. Therefore, the lowest-energy optical bands in LaTiO$_3$ at 1.3\,eV and in YTiO$_3$ at 1.9-2.1\,eV can be associated with a Mott-Hubbard exciton. Alternatively, the lowest-energy bands can possibly be associated with the polaronic or polaron-excitonic origin. However, the modeling of polaron-related features in complex oxides such as CMR materials revealed that the characteristic energies are around 2.4\,eV \cite{Kovaleva_JETP_2002,Kovaleva_PhysB,Kovaleva_PhysB_1}. To clarify this issue, an additional theoretical study exploring the associated energies in LaTiO$_3$ and YTiO$_3$ crystals can be useful.\\        

{\bf Acknowledgement}\\

The author acknowledges Dr. T. Lorentz and Max-Planck Society for making available LaTiO$_3$ single crystal used in the spectroscopic ellipsometry measurements.\\  
 
{\bf Declaration of competing interest}\\

The author declares that she has no known competing financial interests or personal relationships that could have appeared to influence the work reported in this paper.
















\end{document}